# Diffuse neutrino flux measurements with the Baikal-GVD neutrino telescope

**V.M. Aynutdinov[a], V.A. Allakhverdyan[b], A.D. Avrorin[a], A.V. Avrorin[a], Z. Bardačová[c,d], I.A. Belolaptikov[b], E.A. Bondarev[a], I.V. Borina[b], N.M. Budnev[e], V.A. Chadymov[l], A.S. Chepurnov[f], V.Y. Dik[b,g], G.V. Domogatsky[a], A.A. Doroshenko[a], R. Dvornický[c], A.N. Dyachok[e], Zh.-A.M. Dzhilkibaev[a], E. Eckerová[c,d,*], T.V. Elzhov[b], L. Fajt[d], V.N. Fomin[l], A.R. Gafarov[e], K.V. Golubkov[a], N.S. Gorshkov[b], T. I. Gress[e], K.G. Kebkal[h], I.V. Kharuk[a], E.V. Khramov[b], M.M. Kolbin[b], S.O. Koligaev[i], K.V. Konischev[b], A.V. Korobchenko[b], A.P. Koshechkin[a], V.A. Kozhin[f], M.V. Kruglov[b], V.F. Kulepov[j], Y.E. Lemeshev[e], M.B. Milenin[a,†], R.R. Mirgazov[e], D.V. Naumov[b], A.S. Nikolaev[f], D.P. Petukhov[a], E.N. Pliskovsky[b], M.I. Rozanov[k], E.V. Ryabov[e], G.B. Safronov[a], D. Seitova[b,g], B.A. Shaybonov[b], M.D. Shelepov[a], S.D. Shilkin[a], E.V. Shirokov[f], F. Šimkovic[c,d], A.E. Sirenko[b], A.V. Skurikhin[f], A.G. Solovjev[b], M.N. Sorokovikov[b], I. Štekl[d], A.P. Stromakov[a], O.V. Suvorova[a], V.A. Tabolenko[e], B.B. Ulzutuev[b], Y.V. Yablokova[b], D.N. Zaborov[a], S.I. Zavyalov[b], D.Y. Zvezdov[b], N.A. Kosogorov[m,n,s], Y.Y. Kovalev[o,m,n], G.V. Lipunova[p,o], A.V. Plavin[n], D.V. Semikoz[q], and S.V. Troitsky[b,s]**

[a] *Institute for Nuclear Research, Russian Academy of Sciences, Moscow, Russia*

[b] *Joint Institute for Nuclear Research, Dubna, Russia*

[c] *Comenius University, Bratislava, Slovakia*

[d] *Czech Technical University in Prague, Institute of Experimental and Applied Physics, Czech Republic*

[e] *Irkutsk State University, Irkutsk, Russia*

[f] *Skobeltsyn Institute of Nuclear Physics, Moscow State University, Moscow, Russia*

[g] *Institute of Nuclear Physics of the Ministry of Energy of the Republic of Kazakhstan, Almaty, Kazakhstan*

[h] *LATENA, St. Petersburg, Russia*

[i] *INFRAD, Dubna, Russia*

[j] *Nizhny Novgorod State Technical University, Nizhny Novgorod, Russia*

[k] *St. Petersburg State Marine Technical University, St. Petersburg, Russia*

[l] *Moscow, free researcher*

[m] *Astro Space Center of Lebedev Physical Institute, Profsoyuznaya 84/32, 117997 Moscow, Russia*

[n] *Moscow Institute of Physics and Technology, Institutsky per. 9, Dolgoprudny 141700, Russia*

[o] *Max-Planck-Institut fur Radioastronomie, Auf dem Hugel 69, 53121 Bonn, Germany*

[p] *Sternberg Astronomical Institute, Lomonosov Moscow State University, Universitetskii pr. 13, Moscow, 119234 Russia*

[q] *APC, Universite Paris Diderot, CNRS/IN2P3, CEA/IRFU, Sorbonne Paris Cite, 119 75205 Paris, France*

[r] *Physics Department, Lomonosov Moscow State University, 1-2 Leninskie Gory, Moscow 119991, Russia*

[s] *Cahill Center for Astronomy and Astrophysics, MC 249-17 California Institute of Technology, Pasadena, CA 91125, USA*

*E-mail:* eliska.eckerova@fmph.uniba.sk

**Speaker*

*†Deceased*

Baikal-GVD is a next generation, kilometer-scale neutrino telescope currently under construction in Lake Baikal. GVD consists of multi-megaton subarrays (clusters) and is designed for the detection of astrophysical neutrino fluxes at energies from a few TeV up to 100 PeV. The large detector volume and modular design of Baikal-GVD allows for the measurements of the astrophysical diffuse neutrino flux to be performed already at early phases of the array construction. We present here recent results of the measurements on the diffuse cosmic neutrino flux obtained with the Baikal-GVD neutrino telescope using cascade-like events.

## 1. Introduction

Detection of neutrinos with neutrino telescopes is achieved through detecting the Cherenkov radiation emitted by secondary particles produced in neutrino interactions. Charged current (CC) muon neutrino interactions yield long-lived muons that can pass several kilometers through the water or ice, leading to a track signature in the detector. For high-energy muon-like events the accuracy of track reconstruction is typically better than 1°. Neutral current (NC) neutrino interactions and CC interactions of electron and tau neutrinos generally yield cascades - hadronic and electromagnetic showers of charged particles. Directional resolution for cascades is typically a few degrees (for sea- and lake-based experiments). An advantage of the cascade detection channel (over track detection) is its high energy resolution (10–30%) as well as a low atmospheric neutrino background. The cascade channel allows for effective measurement and characterization of the energy-dependent astrophysical neutrino flux. IceCube discovered a diffuse flux of high-energy astrophysical neutrinos in 2013 [1]. The flux, observed using 6 years of IceCube cascade data [2] is consistent with an isotropic single power law model with spectral index $\gamma = 2.53$ and a flux normalization for each neutrino flavor of $\varphi_{astro} = 1.66$ in units of $GeV^{-1}$ $cm^{-2}$ $s^{-1}$ $sr^{-1}$ at $E_0 = 100$ TeV.

The deep underwater neutrino telescope Baikal-GVD (Gigaton Volume Detector) is currently under construction in Lake Baikal [3]. Baikal-GVD is formed by a three-dimensional lattice of optical modules, which consist of photomultiplier tubes housed in transparent pressure-resistant spheres. They are arranged at vertical load-carrying cables to form strings. The telescope has a modular structure and consists of functionally independent clusters - sub-arrays comprising 8 strings. Each cluster is connected to the shore station by an individual electro-optical cable. The first full-scale Baikal-GVD cluster was deployed in April 2016. In 2017–2023, eleven additional clusters were deployed and commissioned, increasing the total number of optical modules to 3456 OMs. The current rate of array deployment is about two clusters per year. One search strategy for high-energy neutrinos with Baikal-GVD is based on the selection of cascade events generated by neutrino interactions in the sensitive volume of the array [4]. Here we discuss the results based on data accumulated in 2018-2021.

## 2. Cascade detection with GVD cluster

The procedure for reconstructing the parameters of high-energy showers - the shower energy, direction, and vertex - is performed in two steps [4]. In the first step, the shower vertex coordinates are reconstructed using the time information from the telescope's triggered photo-sensors. In this case, the shower is assumed to be a point-like source of light. The reconstruction quality can be increased by applying additional event selection criteria based on the limitation of the admissible values for the specially chosen parameters characterizing the events. The vertex reconstruction resolution with this method is about 2 m. In the second step, shower energy and direction are reconstructed by applying the maximum-likelihood method and using the shower coordinates reconstructed in the first step. The values of the variables $\theta$, $\varphi$, and $E_{sh}$ corresponding to the maximum value of the likelihood functional are chosen as the polar and azimuth angles characterizing the direction and the shower energy. The accuracy of cascade energy reconstruction is about 10-30% and the accuracy of direction reconstruction is about 2-4 degree (median value) depending on location and orientation of the cascade.

The search for high-energy neutrinos with a GVD-cluster is based on the selection of cascade events generated by neutrino interactions in the sensitive volume of the array. Performances of the event selection and cascade reconstruction procedures were tested by MC simulation of signal and background events. We used only OM hits with charge Q > 1.5 p.e. Such selection allows for substantial suppression of the noise pulses from water luminescence. After reconstruction of cascade vertex, energy and direction and applying quality cuts, events with a final multiplicity of hit OMs $N_{hit} \geq 8$ were selected as high-energy cascades. The requirement of high hit multiplicity allows substantial suppression of background events from atmospheric muon bundles.

## 3. Data analysis and results

We use Baikal-GVD data collected between April 2018 and March 2022 for the search for astrophysical neutrinos in cascade mode. The telescope was operating in the configuration with 3 clusters in 2018–2019, 5 clusters in 2019–2020, and 7 clusters in 2020–2021, while from April 2021 to March 2022, the telescope consisted of 8 clusters. A sample of $3.49 \times 10^{10}$ events was collected by the basic trigger of the telescope. After applying noise hit suppression procedures, cascade reconstruction and applying cuts on reconstruction quality parameters a sample of 14328 cascades with reconstructed energy $E_{sh}$ > 10 TeV and OM hit multiplicity $N_{hit}$ > 11 was selected.

### 3.1 All-sky analysis

Following the same procedure as in our previous analyses [4], high-energy cascade events with OM hit multiplicity $N_{hit}$ > 19 and reconstructed energy $E_{sh}$ > 70 TeV were selected and additional cuts which suppress events from atmospheric muons were applied [5]. The fraction of background events from atmospheric muons in the selected sample is expected at a level of 50%. As a result, a total of 16 events were selected from the 2018–2021 data [6]. A total of 8.2 events are expected from the simulations of the background (7.4 from atmospheric muons and 0.8 from atmospheric neutrinos) and 5.8 events are expected from the astrophysical best-fit flux derived in [6]. The significance of the excess was estimated to be 2.22σ with the null-cosmic hypothesis rejected at 97.36% confidence level. Background events in the downward going region are dominated by atmospheric muon bundles while atmospheric neutrinos are subdominant by more than an order of magnitude. This analyzed dataset included an event with the energy of the order of 1 PeV. This was the first event with energy of such scale, which was selected from the Baikal-GVD data. The null-cosmic hypothesis is rejected at 99.46% confidence level (2.78σ significance of excess) for such event detection.

### 3.2 Upward-going cascade analysis

Restricting the analysis to upward-going directions allows for effective suppression of the atmospheric muon background, thus improving the neutrino sample purity and enabling the extension of the analysis toward lower energies. Cascadelike events with reconstructed energy $E_{sh}$ > 15 TeV, OM hit multiplicity $N_{hit}$ > 11 and reconstructed zenith angle $\cos\theta < -0.25$ were selected as astrophysical neutrino candidates [6]. A total of 11 events have been selected from the 2018-2021 data sample, while 3.2 atmospheric background events are expected (2.7 from atmospheric conventional and prompt neutrinos and 0.5 events from misreconstructed atmospheric muons). Taking into account the systematic effects, the significance of the excess was estimated to be 3.05σ with the null-cosmic hypothesis rejected at 99.76% C.L. The energy and zenith distributions of the 11 events are shown in Figure 1 together with the distributions obtained by Monte Carlo simulation.

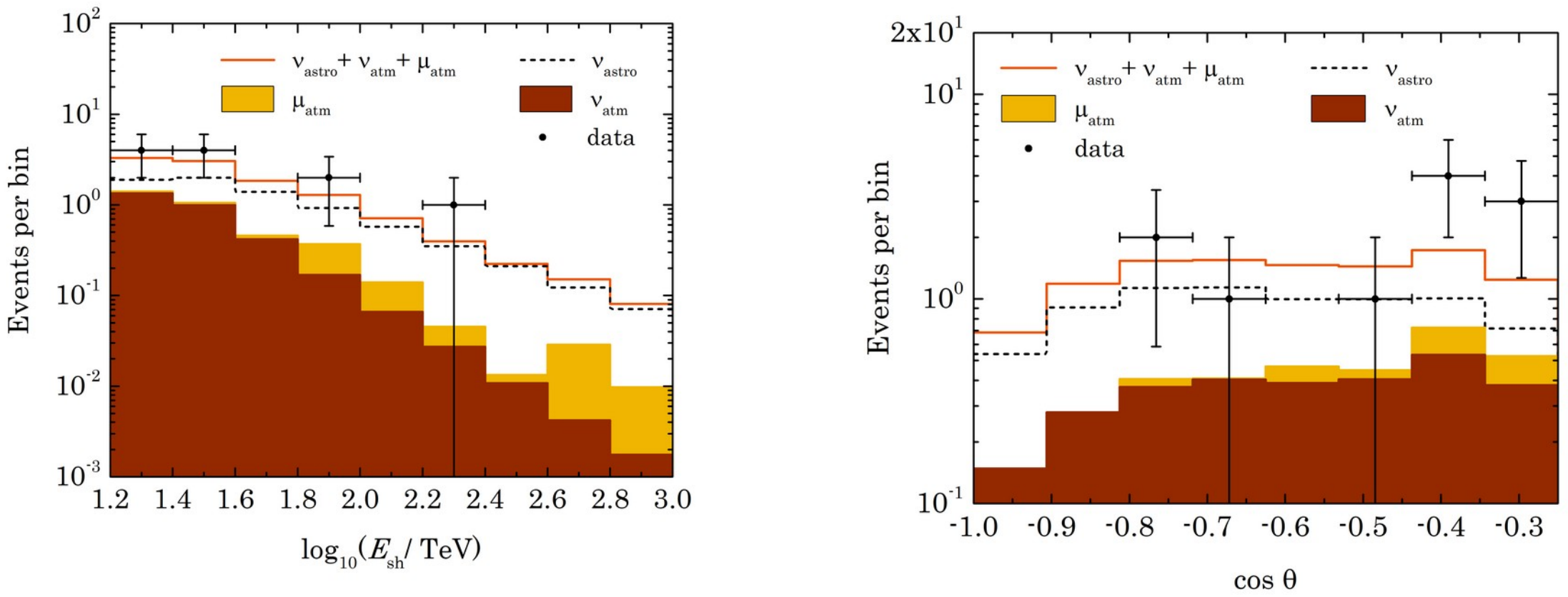

**Figure 1.** Reconstructed cascade energy (left panel) and zenith (right panel) distributions obtained in the upward-going cascade analysis. Black points are data, with statistical uncertainties. The best-fit distribution of astrophysical neutrinos (dashed line), expected distributions from atmospheric muons (yellow) and atmospheric neutrinos (brown) and the sum of the expected signal and background distributions (orange line) are also shown. The atmospheric background histograms are stacked (filled colors).

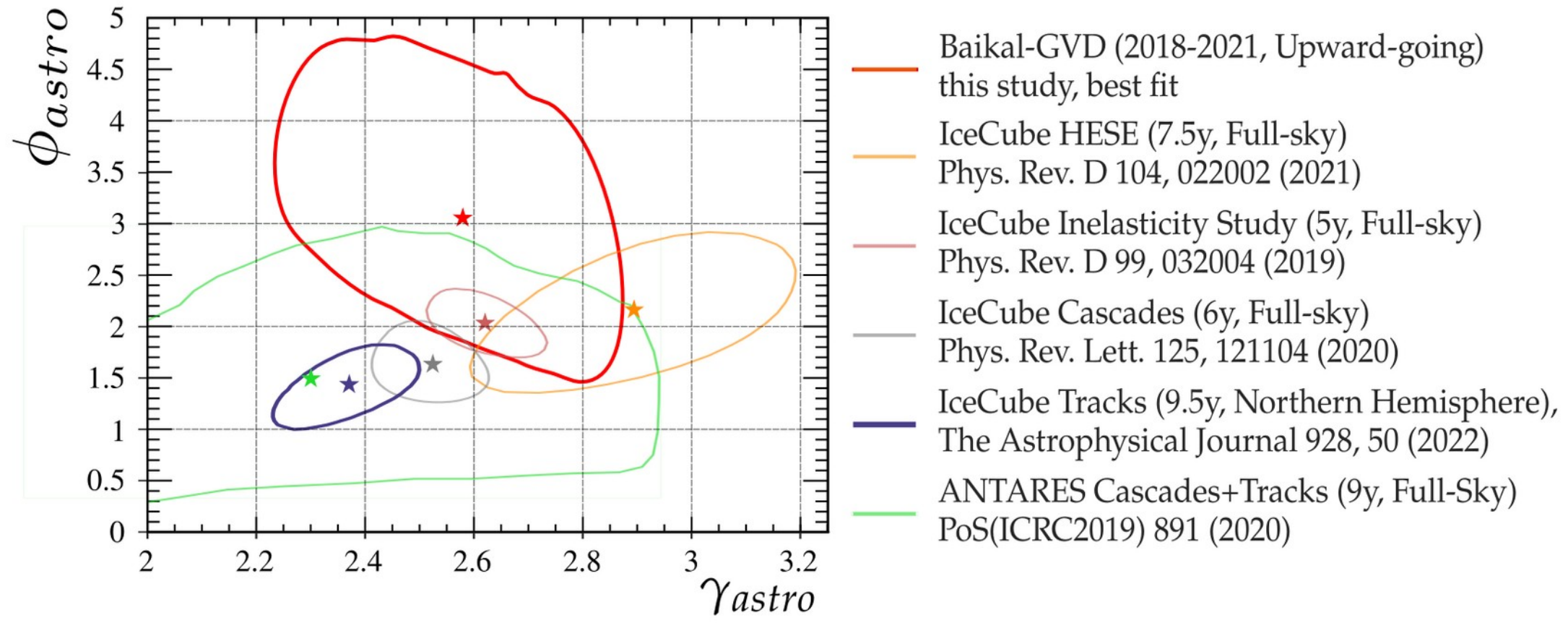

**Figure 2**. The best fit parameters and the contours of the 68% confidence region (red curve) for the single power law hypothesis obtained in the upward-going cascade analysis of the Baikal-GVD data. Other best fits are shown for studies based on high-energy starting events (orange curve) [7], cascadelike events (gray curve) [9], an inelasticity study (purple curve) [10] and track-like events (blue curve) [8] by IceCube and a combined study of tracks and cascades by ANTARES (green curve) [12].

The measured 11 events and the expected number of background events have been analyzed to characterize the diffuse astrophysical neutrino flux. We parametrize the isotropic diffuse astrophysical neutrino flux $\Phi_{astro}$ in the single power law model assuming equal numbers of neutrinos and antineutrinos and equal neutrino flavors at Earth. The model is characterized by spectral index $\gamma_{astro}$ and normalization $\varphi_{astro}$ of the one-flavor neutrino flux in units of GeV$^{-1}$ cm$^{-2}$ s$^{-1}$ sr$^{-1}$. We find the best-fit parameters as following: the spectral index $\gamma_{astro}$ = 2.58 and the flux normalization for each neutrino flavor at $E_0$ = 100 TeV $\varphi_{astro}$ = 3.04. The best-fit parameters and 68% C.L. contours for this cascade analysis together with the results from other neutrino telescopes [7–12] are shown in Figure 2. The Baikal-GVD upward-going neutrino (cascades) measurements are consistent with the IceCube measurements and the ANTARES all-neutrino flavor measurements.

### 3.3 Promising coincident sources

In this section, we scrutinize potential coincident sources associated with the Baikal-GVD cascades, incorporating both Galactic and extragalactic ones [13]. Figure 3 shows the reconstructed sky map positions and the uncertainty regions of the cascade events selected in the all-sky analysis (solid circles) and the upward-going cascade analysis (dashed circles).

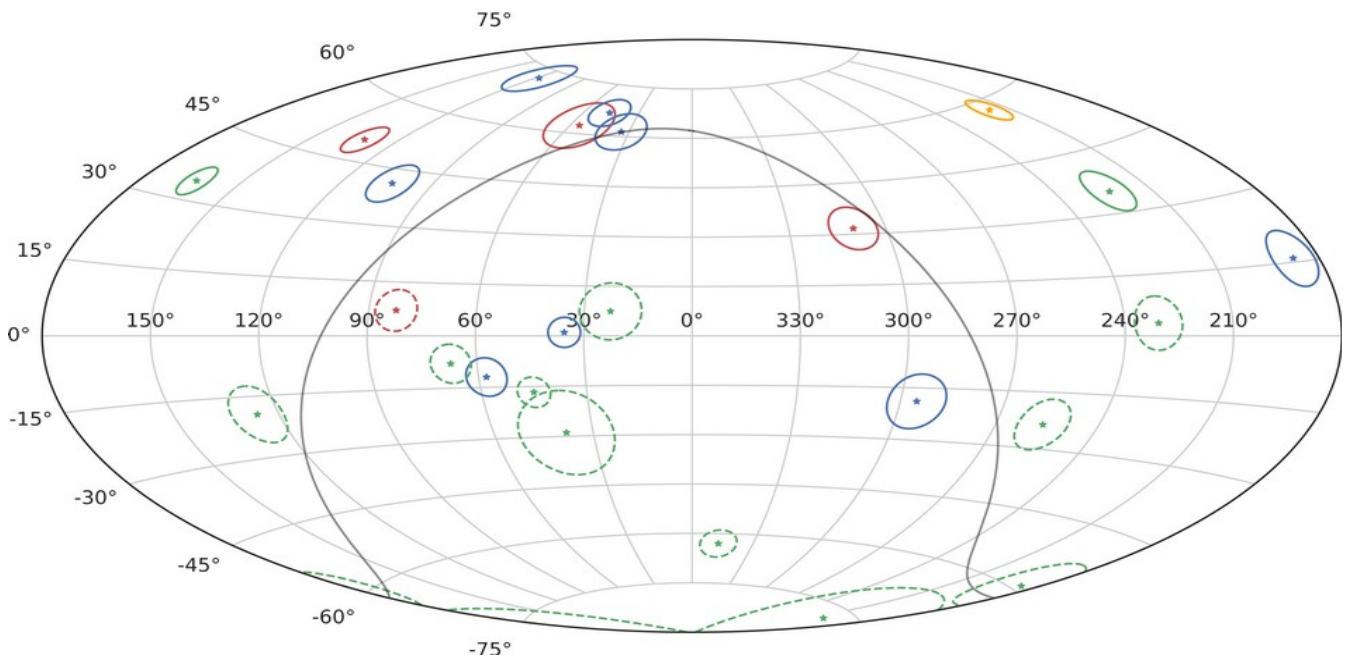

**Figure 3**. The Baikal-GVD high-energy cascade skymap (in equatorial coordinates). The best-fit positions and 90% angular uncertainty regions are shown. Dashed circles show underhorizon events (selected in the upward-going analysis), while solid circles represent events above horizon (selected in the all-sky analysis, excluding upward-going events). Color represents energy of the events: green is below 100 TeV, blue is between 100 TeV and 200 TeV, red is between 200 TeV and 1000 TeV, and orange is above 1 PeV. The Galactic plane is indicated as a gray curve.

Using the Baikal-GVD high-energy cascade sample, it was noted that the error circles (90% CL) of arrival directions of three events intersect, forming a triplet close to the Galactic plane [6]. We estimate the p-value of finding such a triplet on the sky equal to 0.024 (2.26σ ) [13]. The intersection of two of these events contains very well-known Galactic sources of high-energy radiation, LS I +61 303 and Swift J0243.6+6124. LS I +61 303 is a binary system located at 2.6 ± 0.3 kpc from the Solar system. The coordinates for this source are $40^{\circ}.131917$, $61^{\circ}.229333$ and its location is consistent with two cascade events (events GVD190216CA and GVD190604CA). It is known to be variable, with flares coming periodically in a wide range of the electromagnetic spectrum from radio to TeV. The compact object in this system is most probably a pulsar [14]. The orbital period of the binary system is 26.496 d. Additionally, flare brightness modulations are observed in all energy ranges with a longer period ∼ 1659 d. The Baikal-GVD events came at phases 0.3037 and 0.4098 of the orbital period and 0.1403 and 0.2059 of the major period. The latest of the high γ-ray states of LS I +61 303 started in March 2019 and lasted for 96 weeks [15]. Almost in the same position of sky there is the Galactic X-ray pulsar Swift J0243.6+6124 with coordinates are $40^{\circ}.918437$, $61^{\circ}.434377$. The source was discovered in 2017 when a giant flare was taking place in it. Currently, this is the only known pulsating ultraluminous X-ray source (PULX) in the Galaxy. A radio jet was observed from Swift J0243.6+6124 at the time of the X-ray peak in 2017 [16]. In 2019, the source underwent another X-ray outburst, about 6 times dimmer than that of 2017. GVD190216CA happened at its end, and according to [17] the mid of February 2019 is the time of maximal dissipation of the Be-star's decretion disc. The Swift/BAT 15-50 keV daily average light curve shows a flux increase from LS I +61 303 a few days before the neutrino GVD190604CA. Formally, this was the largest flux from LS I +61 303 for the entire observation period since 2005. However, at the time Swift observed the sky area without 'dithering' mode, which indicates a low significance of data. Fermi LAT 0.1-300 GeV data indicates enhanced (about two-fold) radiation at the beginning of June 2019 compared with the long-term period-stacked light curves. At the time of the third event from the triplet, GVD210716C, Swift/BAT, MAXI, or HMXT data show no suspicious features for LS I +61 303 and Swift J0243.6+6124. Thus, existing data does not allow a certain conclusion about the association of any triplet cascade events with high-energy electromagnetic radiation from either of the sources. It needs to be noted that ~ 50% of downward-going neutrino candidates are expected to come from astrophysical objects, while another half are due to background. If both events, GVD190216CA and GVD190604CA, are true neutrinos coming from the same astrophysical source, LS I +61 303 or Swift J0243.6+6124 (or both), this would be the brightest neutrino source in the sky providing a large fraction of the total astrophysical neutrino flux. Intriguingly, the position of one of the events in the Baikal-GVD triplet in the sky coincides with the highest-significance Northern hot spot in the IceCube 7-year sky map [18], though other directions bring higher significances in subsequent similar analyses [19,20]. The corresponding source has not been reported by IceCube at E > 100 TeV, so its observed flux in Baikal-GVD may be an upward fluctuation, and the real flux of this source should be smaller. Notice that, while a continued neutrino emission from the source is not observed by IceCube, a flaring source cannot be ruled out at the moment from a theoretical standpoint [21,22]. At the same time, the arrival of three unrelated high-energy neutrinos from directions close the Galactic plane is also possible and consistent with the recent finding that the IceCube high-energy neutrino flux has a Galactic component [23,24].

Following Plavin et al. [25,26], we select blazars with the brightest radio emission as potential associations with high-energy cascade events [13]. There are four blazars with historical VLBI flux density at 8 GHz above 1 Jy that coincide with the Baikal-GVD high-energy cascade events. Three of them are close to event GVD210409CA. The brightest is 2023+335 (z = 0.22): its average VLBI flux density is 2 Jy at 8 GHz, with a flat or slightly inverted spectrum at GHz frequencies. Moreover, 2023+335 is detected by Fermi LAT in gamma rays [27]. Two others are 2021+317 (z = 0.36) with 1.5 Jy VLBI flux density at 8 GHz and a flat spectrum as well as 2050+364 (z = 0.35) with 1.2 Jy, a steep spectrum, and a lower dominance of parsec-scale structures. The fourth bright blazar is 0529+075 (z = 1.25, [28]) coincident with the event GVD210418CA. Its VLBI flux density is 1.3 Jy at 8 GHz and it has a highly variable spectrum, ranging from falling to inverted. Additionally, there are several coincidences that show evidence of temporal correlation: an ongoing major radio flare when neutrino is detected from the direction of the source. Such behavior is characteristic to neutrino-associated blazars, see [25,26,29]. The most notable is the flaring blazar TXS 0506+056 (z = 0.34, average 0.5 Jy at 8 GHz) coincident with the event GVD210418CA. This association is investigated further in a dedicated paper [30]. Other blazars with hints for temporally coincident flares are 0258−184 (events GVD190523CA and GVD210501CA), and 1935−179 (event GVD200826CA). These objects were highlighted and discussed in [4] together with their radio light curves.

### 3.4 Preliminary results of 2022 data analysis

We use the Baikal-GVD data collected between April 2022 and March 2023 for the search for astrophysical neutrinos in cascade mode. The telescope was operating in the configuration with 10 clusters. A sizeable fraction of these data have not been fully calibrated yet and therefore could not be included in the present analysis at this time. So the analysis is considered preliminary. A sample of $6.3 \times 10^9$ events was collected by the basic trigger of the telescope. After applying noise hit suppression procedures, cascade reconstruction and applying cuts on reconstruction quality parameters a sample of 3226 cascades with reconstructed energy $E_{sh} > 10$ TeV and OM hit multiplicity $N_{hit} > 11$ was selected. All sky analysis and upward-moving cascades analysis were used for the selected data sample. For the all-sky analysis, with the additional requirements $N_{hit} > 19$ and $E_{sh} > 70$ TeV a total of 6 events were selected. As a result, the total number of neutrino candidates selected with the combined 2018-2022 data reached 22 events. A total of 11.3 events are expected from the simulations of the atmospheric background. The significance of the excess was estimated to be 2.74σ (statistical uncertainties only). For the upward-moving cascades analysis 6 events were selected from the 2022 data sample. A total number of 17 neutrino candidates were selected from combined the 2018-2022 data sample. A total of 4.4 events are expected from the simulations of the atmospheric background. The significance of the excess was estimated to be 4.31σ (statistical uncertainties only). The energy and zenith distributions of the 17 events are shown in Figure 4 together with the distributions obtained by Monte Carlo simulation.

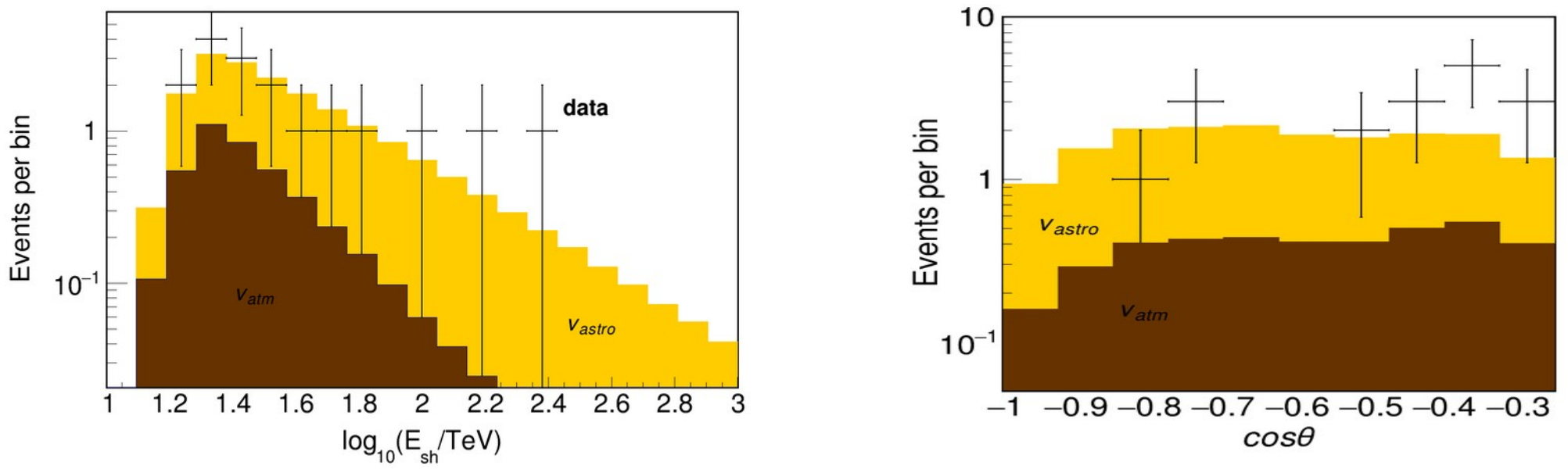

**Figure 4.** Reconstructed cascade energy (left panel) and zenith (right panel) distributions obtained in the upward-going cascade analysis. Black points are data, with statistical uncertainties. The best-fit distribution of astrophysical neutrinos [6] (yellow) and expected distributions from atmospheric neutrinos (brown) are also shown. The atmospheric neutrino background and astrophysical histograms are stacked (filled colors). Due to relatively small number of expected background events from atmospheric muons (0.7 events) they are not shown here.

## Conclusion

We presented the measurements of astrophysical neutrino flux using samples of cascade events collected by the Baikal-GVD in 2018–2021. Using a subsample of upward moving cascades with energy $E_{sh}$ > 15 TeV a total of 11 events have been selected as astrophysical neutrino candidates, while 3.2 atmospheric background events are expected. The significance of the excess over the expected number of atmospheric background events was estimated as 3.05σ. A preliminary analysis of the 2022 data-taking season (data collected between April 2022 and March 2023, upward-moving cascades) has also been presented, which brings the total statistical significance of the diffuse flux detection with upward moving cascades in Baikal-GVD up to 4.31 σ (statistical errors only). We have made a global fit to these neutrino data, fitting the cascade energy distribution, to extract information about the astrophysical neutrino flux. The measured values of an astrophysical power law spectral index of $\gamma_{astro}$ = 2.58 and the flux normalization for each neutrino flavor at $E_0$ = 100 TeV $\varphi_{astro}$ = 3.04 are in good agreement with the previous fits derived in various analyses of the IceCube data and ANTARES data. With these results we, for the first time, confirm the IceCube observation of astrophysical diffuse neutrino flux with 3σ significance. We have discussed promising coincident high-energy astrophysical sources, including Galactic and extragalactic ones. The most notable Galactic sources turned out to be LS I +61 303 and Swift J0243.6+6124, which fall within the 90% uncertainty regions of two events. The extragalactic sources that caught our attention are 2023+335, 2021+317, 2050+364, and 0529+075, which have high values of their average flux densities. We note that TXS 0506+056, 0258−184 and 1935−179 demonstrated temporal coincidences of radio flares and times of neutrino arrival.